\documentclass[prl, groupedaddress, amsmath, amssymb, reprint]{revtex4-1}

\pdfoutput=1

\usepackage{graphicx}
\usepackage{hyperref}

\begin{document}

\title{Plasma Wave Undulator for Laser-Accelerated Electrons}

\author{S. Corde}
\author{K. Ta Phuoc}

\affiliation{Laboratoire d'Optique Appliqu\'ee, ENSTA ParisTech - CNRS UMR7639 - \'Ecole Polytechnique, Chemin de la Huni\`ere, 91761 Palaiseau, France}

\begin{abstract}
Laser-plasma accelerators have become compact sources of ultrashort electron bunches at energies up to the gigaelectronvolt range thanks to the remarkable progresses made over the past decade. A direct application of these electrons bunches is the production of short pulse x-ray radiation sources. In this letter, we study a fully optically driven x-ray source based on the combination of a laser-plasma accelerator and a plasma wave undulator. The longitudinal electric field of a laser-generated plasma wave is used to wiggle electrons transversally. The period of this plasma undulator being equals to the plasma wavelength, tunable photon energies in the ten keV range can be achieved with electron energies in the 100-200 MeV range. Considering a tens terawatt class femtosecond laser system, undulators with a strength parameter $K\sim0.5$ and with about ten periods can be combined with a laser-plasma accelerator, resulting in several $10^{-2}$ emitted x-ray photons per electron.  
\end{abstract}

\maketitle

Laser wake field acceleration (LWFA) has been one of the most developed topic of high intensity laser-plasma interaction during the past ten years. It has lead to number of theoretical and experimental efforts resulting in remarkable results. In LWFA, electrons are accelerated from rest to relativistic energies thanks to strong longitudinal electric field (100's GeV/m) of a large amplitude plasma wave driven in the wake of an intense femtosecond laser pulse. Electron beams currently produced are quasi-monoenergetic \cite{Nature2004Mangles, Nature2004Geddes, Nature2004Faure} with an energy spread down to below the percent \cite{PRL2009Rechatin1}, an energy up to the GeV \cite{NatPhys2006Leemans}, a charge in the 10-100 picocoulomb range, a divergence of a few mrad, a duration in the  few femtoseconds range \cite{NatPhys2010Lundh}, and a normalized emittance in the $\pi.\rm{mm}.\rm{mrad}$ range. 

In the interaction regimes of laser wakefield electron acceleration, there usually occurs the emission of intense and collimated femtosecond x-ray bursts.  Several mechanisms, based on the radiation from relativistic electrons from plasmas, have been studied as sources of femtosecond x-rays:  the non-linear Thomson scattering of a laser beam by plasma electrons \cite{Nature1998Chen, PRL2003TaPhuoc}, the Betatron oscillations of accelerated electrons in the bubble regime \cite{PRL2004Rousse}, the Thomson backscattering of a laser beam by laser-accelerated electrons \cite{PRL2006Schwoerer}, the oscillations of laser-accelerated electrons in a periodical structure of magnets (the conventional meter-scale undulators used in synchrotron facilities) \cite{NatPhys2008Schlenvoigt, NatPhys2009Fuchs}. These high-brightness compact and ultrashort x-ray sources can potentially have many applications in fundamental research as well as in the industrial or medical domains. However, even if these sources can already cover a spectral range from a few eV (conventional undulators) to a few 100 keV (Thomson backscattering) there exists a gap in the ten keV range where none of these sources is efficient.  Only the $K_\alpha$ source emits in this energy range but it is a fully isotropic radiation and has a duration of a few hundreds femtosecond. The production of a radiation beam with a few femtosecond duration at this energy is therefore particularly important for applications. Indeed, most of the applications require a source with an \aa ngstr\"om wavelength and a femtosecond duration, respectively the typical interatomic distances and time scales of fundamental processes in matter. 

In this letter, we discuss the possibility to produce a fully optically driven femtosecond and tunable x-ray source in the ten keV range by using a laser-generated plasma wave as an undulator for a laser-accelerated electron bunch. The concept of the plasma wave undulator was originally proposed in the context of realization of compact short wavelength free electron lasers from conventional accelerators \cite{PROC1987Joshi, IEEE1987Joshi}. Indeed, such an undulator has a very small period compared to conventional undulators \cite{Clarke} and allows to produce short wavelength free electron radiation with moderate electron energies. Further studies on the plasma wave undulator were performed a few years later \cite{RSI1990Williams, PROC1990Williams, PS1990Fedele, IEEE1993Williams}, discussing in more details the classical spontaneous emission. Electron trajectories as well as the incoherent radiation they emit were analytically or numerically calculated using ideal linear plasma waves. In the following, we will study the utility of the plasma wave undulator for laser-plasma accelerators. The interest of this study is motivated by the active research on femtosecond x-ray sources and by the fact that laser-plasma accelerators have become efficient and reliable at the 100 MeV energy level. We will show that, at these electron energies, high brightness and tunable undulator type radiation can be produced in the relevant ten keV range. Our description does not rely on ideal linear plasma wave. Instead, we first propose a concise analytical analysis of the nonlinear one-dimensional plasma wave undulator for estimation of the radiation properties, and then fully describe the realistic plasma wave generation using quasi-static particle code simulations \cite{PRE1996Mora}. These realistic plasma waves are then used to calculate numerically electron trajectories and the emitted x-ray radiation. The consideration of realistic plasma waves is particularly important since for the considered amplitudes they become non linear, with possibly damping and breaking of the plasma wave, and strong modification of the driving laser pulse along its propagation in the plasma. All these effects are taken into account in the used quasi-static particle code WAKE \cite{PRE1996Mora}.

A plasma wave corresponds to the coupled propagation of the electron density $n_e$ and the scalar/electric  potential $V$. For non-linear multi-dimensional plasma waves, the charge current $\vec{j}_e$ and the magnetic field $\vec{B}$ also appear in their description. Several routes exist to produce plasma waves. They have been explored in order to produce intense electric fields for electron acceleration in plasmas. Large amplitude plasma waves can be generated through the beating of two copropagating long laser pulses satisfying the condition $\omega_2-\omega_1=\omega_p$ [$\omega_p=(n_ee^2/m\epsilon_0)^{1/2}$ is the plasma frequency, $\omega_1$ and $\omega_2$ the laser frequencies of each laser pulse], the self-modulation instability [a single long laser pulse breaks in a pulse train of duration $\tau\sim\lambda_p/c$, where $\lambda_p=2\pi c/\omega_p$ is the plasma wavelength] and the resonant case where the laser pulse has a duration $\tau\sim\lambda_p/c$. These methods have been demonstrated to be efficient for the generation of electric fields $\vec{E}=-\vec{\nabla}V$ in the 100's GeV/m range (it has to be compared with the highest achievable electric field by conventional approaches, which is in the 10's MeV/m range). In the following, we will consider the generation of plasma waves by a resonant excitation.

\begin{figure}
   \includegraphics[width=8.5cm]{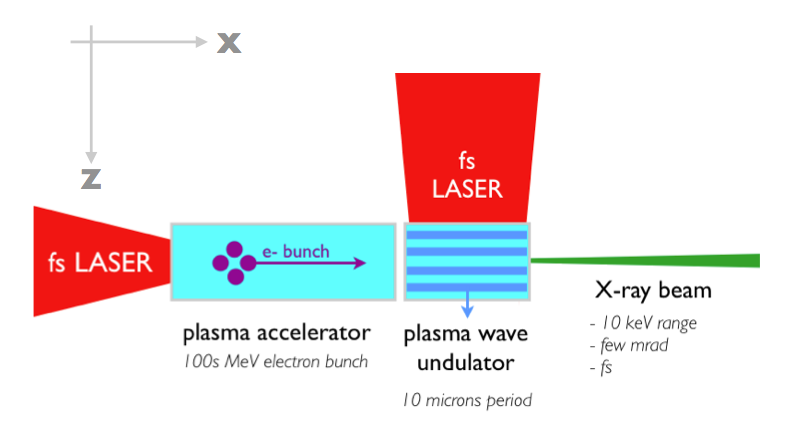}
\caption{Principle of the plasma undulator source combined with a laser-plasma accelerator. An electron bunch is created and accelerated at the interaction of an intense femtosecond laser with a millimeter scale plasma. Once accelerated electrons propagate and are wiggled in a plasma wave. Their relativistic oscillation motion results in the emission of an x-ray beam.}
\label{fig1}
\end{figure}
The schematic of the x-ray source studied is presented in Fig. \ref{fig1}. Electrons are first accelerated to relativistic energies in a laser-plasma accelerator (along the $\hat{x}$ direction) and then propagate in a plasma wave traveling in the $\hat{z}$ direction. As they propagate in the plasma wave, they experience successively an electric field directed toward the $\hat{z}$ positive values and toward the $\hat{z}$ negative values. As a consequence they oscillate in the $\hat{z}$ direction with a significant transverse acceleration. Since both the plasma wave and the electron bunch travel at almost the speed of light, the period of the electron motion is equal to the plasma wave period, i.e. approximately the plasma wavelength (which is $10.6\:\mu$m for an electron density of $10^{19}\:\rm{cm}^{-3}$). The result of this relativistic oscillatory motion is the emission of an x-ray beam by the electron bunch. Experimentally this can be realized by using two synchronized laser pulses. The first laser pulse, directed along the $\hat{x}$ direction, is used to create the plasma accelerator, producing a relativistic electron bunch. The second laser pulse, propagating in the transverse direction $\hat{z}$, used to create the plasma wave undulator.

\bigskip
The radiation produced in the plasma undulator is a relativistic moving charge radiation and its features depend on the electron motion. In order to derive estimations for the electron orbits driven by a plasma wave undulator, we consider a one dimensional plasma wave described by the normalized electric potential $\phi(\xi)=eV/mc^2$ only depending on the comoving variable $\xi=z-v_pt$ (quasi-static approximation), $v_p$ being the plasma wave phase velocity and supposedly equal to the laser group velocity.

We assume an electron from a laser-plasma accelerator with an initial energy $\mathcal{E}=\gamma_imc^2$ ($\gamma_i$ is the initial relativistic factor) and a velocity directed along the $\hat{x}$ axis. The electron enters transversally in the one-dimensional (1D) plasma wave potential $\phi(\xi)$. The equation of the test electron motion is:
\begin{equation}
\frac{d\vec{p}}{dt}=\vec{\nabla}\phi,
\end{equation}
where all quantities are normalized by the choice $m=c=e=\omega_p=1$. The Hamiltonian describing the dynamic of the test electron in the plasma wave is: 
\begin{equation}
\mathcal{H}(\vec{r},\vec{P},t)=\gamma-\phi=\sqrt{1+(\vec{P}+\vec{a})^2}-\phi,
\end{equation}
where $\vec{P}=\vec{p}-\vec{a}$ is the canonical momentum. In our case the canonical and kinematic momentum $\vec{p}$ are equal because the vector potential $\vec{a}=0$. $\mathcal{H}$ does not depend on the $x$ and $y$ coordinates. This implies the conservation of $p_x$ and $p_y$. Furthermore it depends on $t$ and $z$ only through $\xi=z-\beta_pt\simeq z-t$. This gives the following constant of motion:
\begin{equation}
\gamma-\phi-p_z=\textrm{const}.
\end{equation}
Using the initial conditions for the test electron: $x=y=z=p_y=p_z=0$, $p_x=\sqrt{\gamma_i^2-1}$, $\phi=0$, the above constants of motion and the relation $\gamma^2=1+p^2$, we obtain:
\begin{equation}
\label{eqpz}
p_z=-\phi+\frac{\phi^2}{2\gamma_i}+o(\frac{\phi^2}{\gamma_i}),
\end{equation}
where we develop expressions using the approximation $\phi\ll\gamma_i$ which is always satisfied (the normalized potential cannot exceed a few units in realistic plasma waves, whereas $\gamma_i$ is superior to 100 for our case).

The electron motion consists of a drift in the $\hat{x}$ direction combined with an oscillation in the transverse direction $\hat{z}$. The plasma wave $\phi(\xi)$ acts as an undulator with a period $\lambda_u\simeq\lambda_p$. This period depends on the plasma density and can be controlled. In practical units, it is given by $\lambda_u[\mu\textrm{m}]=3.34\times10^{10}/\sqrt{n_e[\textrm{cm}^{-3}]}$. A second relevant parameter characterizing the plasma undulator is its strength parameter $K$. It is defined as the product of the electron energy $\gamma$ by the maximum angle of the trajectory $\psi$ (with respect to the $\hat{x}$ axis): $K=\gamma\psi$. Because $\psi=dz/dx{_{\text{max}}}\simeq dz/dt{_{\text{max}}}=p_z{_{\text{max}}}/\gamma$, from Eq. (\ref{eqpz}) the strength parameter $K$ of the plasma wave undulator reads
\begin{equation}
K=\gamma\psi=p_z{_{\text{max}}}=\phi_0,
\end{equation}
where $\phi_0$ is the amplitude of the plasma wave potential $\phi(\xi)$. In the linear limit $a_0^2\ll1$, the plasma wave reads $\phi(\xi)=\sqrt{\pi}a_0^2(\omega_p\tau/4)\exp(-\omega_p^2\tau^2/4)\sin(k_p\xi)$, for a gaussian laser pulse with duration $\tau$ at $1/e^2$ in intensity, and if we approach the non-linear regime, we can reach $K=\phi_0\lesssim1$.

The radiation emitted by an electron traveling in the plasma wave undulator is a moving charge emission described by the general expression giving the radiation emitted by an electron, in a direction of observation $\vec{n}$, as a function of its position, velocity and acceleration along the trajectory \cite{Jackson}:
\begin{equation}
\label{rayonnement}
\frac{d^2I}{d\omega d\Omega} = \frac{e^2}{16 \pi ^3 \epsilon_0 c} \left
|\int_{-\infty}^{+\infty} e^{i \omega (t-\vec{n}. \vec{r}(t)/c)}
\frac{\vec{n} \times \left [ (\vec{n}-\vec{\beta}) \times
\dot{\vec{\beta}} \right ]}{(1-\vec{\beta}.\vec{n})^2} dt \right
|^2.
\end{equation}
This expression represents the energy radiated within a spectral band $d\omega$ centered on the frequency $\omega$ and a solid angle $d\Omega$ centered on the direction of observation $\vec{n}$. Here, $\vec{r}(t)$ is the electron position at time $t$, $\vec{\beta}$ is the velocity of the electron normalized to the speed of light $c$, and $\dot{\vec{\beta}}=d\vec{\beta}/dt$ is the usual acceleration divided by $c$. The analysis of this general expression results in analytical expressions describing the radiation features as a function of the parameters $\gamma_i$, $\lambda_u$ and $K$. In the laboratory frame, the radiation emitted by a relativistic electron at a given time is directed in the direction of its velocity $\vec{v}$, within a cone of typical angle $1/\gamma$. The angular distribution of the radiation depends on the electron orbit. For a trajectory confined in the plane $(\hat{x},\hat{z})$, which is the case for a 1D plasma wave, the radiation is emitted in a cone of half angle $\theta = 1/\gamma$ for $K<1$, while for $K>1$ the angular distribution is increased in the direction of the oscillation  $\hat{z}$ with a half angle of $K/\gamma$. The electron motion being periodic with a spatial period $\lambda_u$, the spectrum of the radiation consists on the fundamental wavelength $\lambda=\frac{\lambda_u}{2\gamma^2}(1+K^2/2+\gamma^2\theta^2)$ and its harmonics, where $\theta$ is the observation angle with respect to $\hat{x}$. For the undulator limit $K\ll1$, i.e small transverse amplitudes of oscillation, the spectrum contains only the fundamental wavelength, and harmonics start to appear when $K\rightarrow1$. In the wiggler limit $K\gg1$, the spectrum contains many closely spaced harmonics extending up to a critical wavelength $\lambda_c=\frac{2\lambda_u}{3K\gamma^2}$. For the case of the plasma wave $K\lesssim1$ and the radiation is produced in the undulator regime at the fundamental wavelength and possibly a few harmonics. The spectral thickness of the fundamental wavelength and its harmonics depend on the number of oscillation periods $N$. It is given by $\Delta \lambda_n/\lambda_n=1/(nN)$ where the index $n$ indicates the harmonic order. Note that the spectrum is broadened with a decreasing energy when observed off axis. The total number of emitted photons depends on the electron energy $\gamma$, the $K$ parameter and the number of oscillations $N$. Over one oscillation period and for $K\ll1$, it is given by $N_{X}= \frac{2\pi}{3}\alpha K^2$ (at the mean energy $\langle\hbar\omega\rangle=\gamma^2hc/\lambda_u$) \cite{Jackson} where $\alpha=e^2/4\pi \epsilon_0\hbar c$ is the fine structure constant. Finally, the duration of the emitted radiation pulse is sensibly equal to the duration of the electron bunch. In the case of laser-plasma acceleration, the duration of the electron bunch can be as small as a few femtoseconds \cite{NatPhys2010Lundh}.

For example, if we consider a 150 MeV electron traveling in a plasma wave undulator with $K=0.5$ and a period $\lambda_u=10$ $\mu$m the produced radiation has an on-axis fundamental energy $E_X \simeq 21$ keV. The half angle of the x-ray beam divergence is $\theta \simeq 3$ mrad and the expected number of photons produced per electron and per period is $\simeq 4\times 10^{-3}$. Assuming $10^8$ electrons and 10 periods the total number of photons is $\simeq 4\times10^6$. This photon energy range is not covered by the existing laser produced x-ray sources. In addition with the present methods of wakefield acceleration, the electron energy can typically be tuned from 50 to 200 MeV \cite{Nature2006Faure}, which corresponds to x-ray energy from $\sim 2$ to 37 keV. 

\bigskip
The above description assumes an idealized situation where the plasma wave is 1D and the electron motion is strictly periodic. A more precise description based on numerical simulations is presented in the following.

The generation of the plasma wave by a resonant excitation has been calculated using the Wake code \cite{PRE1996Mora}. Wake is a two dimensional, moving window particle code describing laser-plasma interaction in the short pulse relativistic regime. It allows to calculate the propagation of a relativistic laser pulse in an underdense plasma and its wakefields. To speed up the calculation time compared to PIC (particle in cell) simulations, Wake assumes that the laser pulse does not evolve significantly during the electron transit time through the laser pulse (quasi-static approximation), and it uses averaged equation of motions for the particles (average over the laser period) so that it is only necessary to resolve the plasma time scale, and not the laser time scale. As a consequence, the code can not describe high energy electrons moving in the forward direction with $\gamma>\omega_0/\omega_p$, where $\omega_0$ is the laser frequency.

The wakefield maps obtained from Wake are then used to calculate the orbits of test electrons in the plasma wave undulator. From the electron trajectories, we compute the x-ray radiation properties using Eq. (\ref{rayonnement}).

\begin{figure}
   \includegraphics[width=8.5cm]{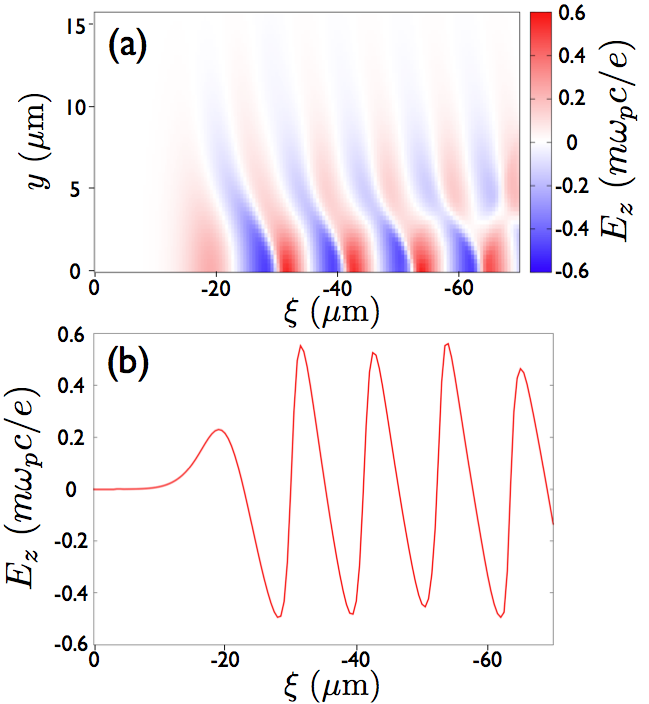}
\caption{The map of the field $E_z$ obtained from the Wake code is displayed in (a). It corresponds to the wakefield after 0.5 mm of propagation of the laser pulse in the gas jet of density $n_e=10^{19}\textrm{cm}^{-3}$. The profile $E_z(\xi)$ at $y=0$ is shown in (b).}
\label{fig2}
\end{figure}
We have considered a laser pulse, propagating in the $\hat{z}$ direction, focused on a gas jet with a density of $n_e=10^{19}\textrm{cm}^{-3}$. The focal spot is chosen to be elliptical, with a waist larger in the $\hat{x}$ direction than in the  $\hat{y}$ direction in order to increase the overlap of electrons with the plasma wave. The laser waist equals $11\:\mu\textrm{m}$ in the $\hat{y}$ direction, the pulse duration is $30\:\textrm{fs}$ at $1/e^2$ in intensity, and the amplitude of the normalized vector potential of the laser pulse is $a_0=1.2$. After $0.5\:\textrm{mm}$ of propagation in the plasma, the Wake simulation [run in the two dimensions $(\hat{z},\hat{y})$] shows that the amplitude of the plasma wave attains $\phi_0\simeq0.5$ , as displayed on Fig. \ref{fig2}. The maps of the wake fields $E_z$, $E_y$ and $B_x$, after $0.5\:\textrm{mm}$ of propagation in the plasma, characterize the plasma wave undulator. To take into account the finite dimension of the laser pulse and of the wake fields in the $\hat{x}$ direction, we apply a gaussian profile with a waist of $44\:\mu\textrm{m}$ to the wake fields for this direction.

We have injected an electron beam of energy $\mathcal{E}=150\:\textrm{MeV}$ and of transverse size $7\:\mu\textrm{m}$, propagating in the $\hat{x}$ direction, in the plasma wave undulator characterized by the fields obtained from Wake, and we have calculated the produced radiation using Eq. (\ref{rayonnement}). The electrons emit a narrow x-ray beam directed in the $\hat{x}$ direction, with a divergence of a few mrad. 
\begin{figure}
   \includegraphics[width=8.5cm]{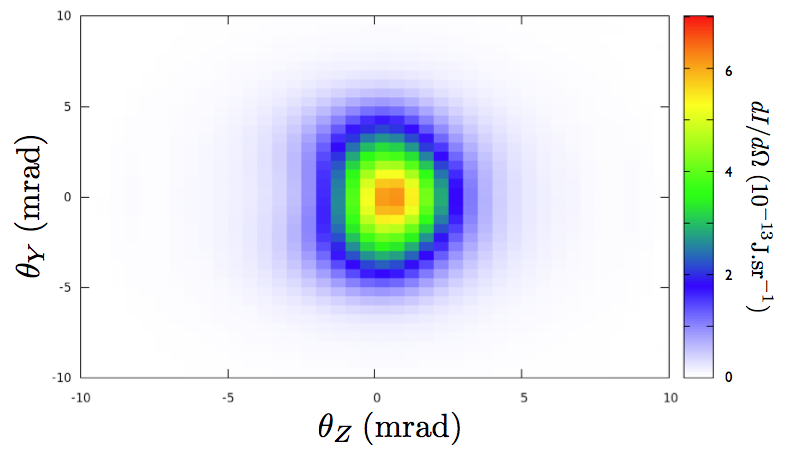}
\caption{x-ray angular profile of the plasma wave undulator radiation, in J/sr and per electron.}
\label{fig3}
\end{figure}
The angular profile of the radiation is shown on Fig. \ref{fig3}. 
\begin{figure}
   \includegraphics[width=8.5cm]{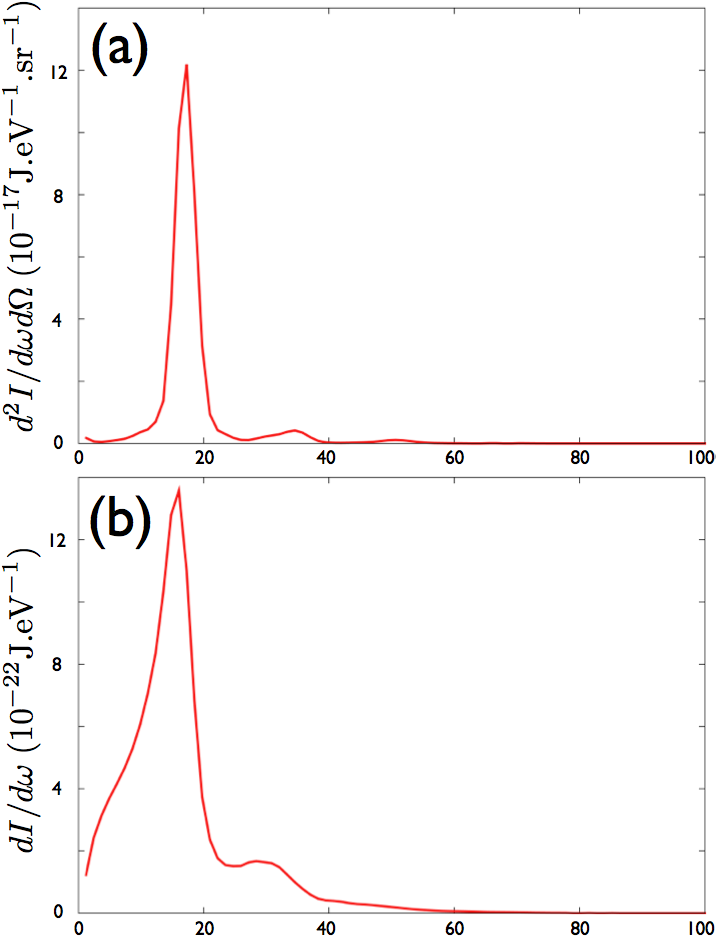}
\caption{The on-axis spectrum ($\theta_Z=\theta_Y=0$) of the plasma wave undulator radiation is displayed in (a), in J/eV/sr and per electron. The integrated spectrum over angles is shown in (b), in J/eV and per electron.}
\label{fig4}
\end{figure}
Figure \ref{fig4} displays the spectra of the radiation emitted on axis as well as integrated over the angular distribution. It is in agreement with the properties of the undulator radiation, with an angle-dependent wavelength and a narrow bandwidth at a given angle of observation. The central photon energy and the full width at half maximum bandwidth of the on-axis spectrum are respectively $17\:$keV and 22\%. The total number of emitted photons is $10^{-2}$ per electron. The peak brightness associated to the on-axis spectrum of Fig. \ref{fig4} is on the order of $10^{20}-10^{21}$ photons/0.1\%bandwidth/s/mm$^2$/mrad$^2$, considering $10^9$ electrons, a source size of 5-10 microns and a duration of a few femtoseconds. These results show good agreement with the estimations obtained with the analytical 1D study.

\bigskip
In conclusion, the combination of a laser-plasma accelerator with a plasma undulator offers the possibility to produce a compact tunable source of ultrashort x-ray radiation in the ten keV range. This source would be tunable in a wide spectral range thanks to both the tunability of the undulator period with the plasma density and the electron energy. This x-ray source exhibits the properties of undulator radiation: narrow bandwidth spectrum and high brightness.

The plasma wave undulator approach presents similarities with the Thomson scattering, and their advantages and disadvantages will be briefly discussed. In the Thomson scattering geometry a laser pulse scatters off the electron beam from the laser-plasma accelerator. Electrons oscillate in the laser pulse and emit radiation that can rapidly attain very high photon energies thanks to two successive Doppler shifts. The oscillation period is given by $\lambda_u\simeq\lambda_0/(1-\beta\cos\theta)$, where $\lambda_0$ is the laser wavelength and $\theta$ the angle between the laser and the electron beam directions (co-propagating for $\theta=0$), and the strength parameter of the oscillation is $K=a_0$ where $a_0$ is the normalized amplitude of the laser pulse vector potential. If we consider the highest quality electron beams from a laser-plasma accelerator produced to date, their energy is in the hundred MeV range. At these energies, the Thomson scattering is more appropriate to produce energetic photons, in the 100 keV or MeV range. Indeed, the production of x-ray radiation at 10 keV using electrons at 200 MeV requires a co-propagating geometry where the laser and the electron beam make about a 10 degrees angle. This complicates the experimental setup due to the positioning of the focusing optics. In addition, the laser scattering off the electron beam has to travel into the few millimeter gas jet, can diffract and lose energy. Both the Thomson scattering and plasma wave undulator sources can have a strength parameter $K$ of the same order of magnitude, $\lesssim1$, but it is easier to reach a higher number of oscillation periods with the Thomson scattering, and so a higher number of photons and brightness. 

The plasma wave undulator appears to be an interesting scheme because its period in the ten microns range is adapted to the electron energies which are currently produced. It allows to reach the 10 keV range using electrons at a few hundreds MeV. In addition, the undulator period and therefore the energy of the radiation can be easily tuned by varying the plasma density, keeping the electron beam constant. Even if the two schemes can therefore be used to access the ten keV range and produce femtosecond x-ray radiation, the Thomson scattering looks more appropriate to access higher energies, the hundred keV range, while the plasma wave undulator is suitable to access the ten keV range.

In this letter we have chosen to discuss the properties of this source as it could be produced with 10-100 TW short pulse laser systems. The numerical simulations, in agreement with analytical predictions, show that a radiation beam, collimated within a few mrad, at energies in the ten keV range and with $\sim 10^{-2}$ photons / electron can be produced. Assuming electron bunches with 60 pC charge, the source can deliver $\sim10^6$ photons / pulse. The source size is equal to the dimension of the electron bunch within the undulator. This depends on the distance between the electron source and the undulator. Typically, a source size on the order of 5 $\mu$m can be expected. The flux of this source is not as high as the betatron source or the $K_\alpha$ source, however this source combines the advantages of narrow band spectrum and tunability, the collimation of a few mrad and the ultra-short pulse duration of a few femtoseconds.
With more laser energy, a major improvement would be to increase the number of periods of the plasma wave undulator, by focusing on a larger waist in the direction of propagation of electrons $\hat{x}$, and by generating a plasma wave consisting of many plasma periods. Assuming a plasma wave undulator with 25 periods and with $K \simeq 0.5$ the source could reach the level of $\sim 10^{-1}$ photons / electron with an on-axis spectral bandwidth of a few percents. The use of future high repetition rate laser systems would allow to improve the averaged brightness of this source.

\bigskip
We acknowledge Professor P. Mora for providing the Wake code.

\bibliographystyle{apsrev4-1}
\bibliography{sebastiencorde}

\begin{thebibliography}{22}%
\makeatletter
\providecommand \@ifxundefined [1]{%
 \@ifx{#1\undefined}
}%
\providecommand \@ifnum [1]{%
 \ifnum #1\expandafter \@firstoftwo
 \else \expandafter \@secondoftwo
 \fi
}%
\providecommand \@ifx [1]{%
 \ifx #1\expandafter \@firstoftwo
 \else \expandafter \@secondoftwo
 \fi
}%
\providecommand \natexlab [1]{#1}%
\providecommand \enquote  [1]{``#1''}%
\providecommand \bibnamefont  [1]{#1}%
\providecommand \bibfnamefont [1]{#1}%
\providecommand \citenamefont [1]{#1}%
\providecommand \href@noop [0]{\@secondoftwo}%
\providecommand \href [0]{\begingroup \@sanitize@url \@href}%
\providecommand \@href[1]{\@@startlink{#1}\@@href}%
\providecommand \@@href[1]{\endgroup#1\@@endlink}%
\providecommand \@sanitize@url [0]{\catcode `\\12\catcode `\$12\catcode
  `\&12\catcode `\#12\catcode `\^12\catcode `\_12\catcode `\%12\relax}%
\providecommand \@@startlink[1]{}%
\providecommand \@@endlink[0]{}%
\providecommand \url  [0]{\begingroup\@sanitize@url \@url }%
\providecommand \@url [1]{\endgroup\@href {#1}{\urlprefix }}%
\providecommand \urlprefix  [0]{URL }%
\providecommand \Eprint [0]{\href }%
\@ifxundefined \urlstyle {%
  \providecommand \doi  [0]{\begingroup \@sanitize@url \@doi}%
  \providecommand \@doi [1]{\endgroup \@@startlink {\doibase
  #1}doi:\discretionary {}{}{}#1\@@endlink }%
}{%
  \providecommand \doi  [0]{doi:\discretionary{}{}{}\begingroup
  \urlstyle{rm}\Url }%
}%
\providecommand \doibase [0]{http://dx.doi.org/}%
\providecommand \Doi [0]{\begingroup \@sanitize@url \@Doi }%
\providecommand \@Doi  [1]{\endgroup\@@startlink{\doibase#1}\@@Doi}%
\providecommand \@@Doi [1]{#1\@@endlink}%
\providecommand \selectlanguage [0]{\@gobble}%
\providecommand \bibinfo  [0]{\@secondoftwo}%
\providecommand \bibfield  [0]{\@secondoftwo}%
\providecommand \translation [1]{[#1]}%
\providecommand \BibitemOpen [0]{}%
\providecommand \bibitemStop [0]{}%
\providecommand \bibitemNoStop [0]{.\EOS\space}%
\providecommand \EOS [0]{\spacefactor3000\relax}%
\providecommand \BibitemShut  [1]{\csname bibitem#1\endcsname}%
\bibitem [{\citenamefont {Mangles}\ \emph {et~al.}(2004)\citenamefont
  {Mangles}, \citenamefont {Murphy}, \citenamefont {Najmudin}, \citenamefont
  {Thomas}, \citenamefont {Collier}, \citenamefont {Dangor}, \citenamefont
  {Divall}, \citenamefont {Foster}, \citenamefont {Gallacher}, \citenamefont
  {Hooker}, \citenamefont {Jaroszynski}, \citenamefont {Langley}, \citenamefont
  {Mori}, \citenamefont {Norreys}, \citenamefont {Tsung}, \citenamefont
  {Viskup}, \citenamefont {Walton},\ and\ \citenamefont
  {Krushelnick}}]{Nature2004Mangles}%
  \BibitemOpen
  \bibfield  {author} {\bibinfo {author} {\bibfnamefont {S.~P.~D.}\
  \bibnamefont {Mangles}}, \bibinfo {author} {\bibfnamefont {C.~D.}\
  \bibnamefont {Murphy}}, \bibinfo {author} {\bibfnamefont {Z.}~\bibnamefont
  {Najmudin}}, \bibinfo {author} {\bibfnamefont {A.~G.~R.}\ \bibnamefont
  {Thomas}}, \bibinfo {author} {\bibfnamefont {J.~L.}\ \bibnamefont {Collier}},
  \bibinfo {author} {\bibfnamefont {A.~E.}\ \bibnamefont {Dangor}}, \bibinfo
  {author} {\bibfnamefont {E.~J.}\ \bibnamefont {Divall}}, \bibinfo {author}
  {\bibfnamefont {P.~S.}\ \bibnamefont {Foster}}, \bibinfo {author}
  {\bibfnamefont {J.~G.}\ \bibnamefont {Gallacher}}, \bibinfo {author}
  {\bibfnamefont {C.~J.}\ \bibnamefont {Hooker}}, \bibinfo {author}
  {\bibfnamefont {D.~A.}\ \bibnamefont {Jaroszynski}}, \bibinfo {author}
  {\bibfnamefont {A.~J.}\ \bibnamefont {Langley}}, \bibinfo {author}
  {\bibfnamefont {W.~B.}\ \bibnamefont {Mori}}, \bibinfo {author}
  {\bibfnamefont {P.~A.}\ \bibnamefont {Norreys}}, \bibinfo {author}
  {\bibfnamefont {F.~S.}\ \bibnamefont {Tsung}}, \bibinfo {author}
  {\bibfnamefont {R.}~\bibnamefont {Viskup}}, \bibinfo {author} {\bibfnamefont
  {B.~R.}\ \bibnamefont {Walton}}, \ and\ \bibinfo {author} {\bibfnamefont
  {K.}~\bibnamefont {Krushelnick}},\ }\Doi {10.1038/nature02939} {\bibfield
  {journal} {\bibinfo  {journal} {Nature (London)},\ }\textbf {\bibinfo
  {volume} {431}},\ \bibinfo {pages} {535} (\bibinfo {year}
  {2004})}\BibitemShut {NoStop}%
\bibitem [{\citenamefont {Geddes}\ \emph {et~al.}(2004)\citenamefont {Geddes},
  \citenamefont {Toth}, \citenamefont {van Tilborg}, \citenamefont {Esarey},
  \citenamefont {Schroeder}, \citenamefont {Bruhwiler}, \citenamefont {Nieter},
  \citenamefont {Cary},\ and\ \citenamefont {Leemans}}]{Nature2004Geddes}%
  \BibitemOpen
  \bibfield  {author} {\bibinfo {author} {\bibfnamefont {C.~G.~R.}\
  \bibnamefont {Geddes}}, \bibinfo {author} {\bibfnamefont {C.}~\bibnamefont
  {Toth}}, \bibinfo {author} {\bibfnamefont {J.}~\bibnamefont {van Tilborg}},
  \bibinfo {author} {\bibfnamefont {E.}~\bibnamefont {Esarey}}, \bibinfo
  {author} {\bibfnamefont {C.~B.}\ \bibnamefont {Schroeder}}, \bibinfo {author}
  {\bibfnamefont {D.}~\bibnamefont {Bruhwiler}}, \bibinfo {author}
  {\bibfnamefont {C.}~\bibnamefont {Nieter}}, \bibinfo {author} {\bibfnamefont
  {J.}~\bibnamefont {Cary}}, \ and\ \bibinfo {author} {\bibfnamefont {W.~P.}\
  \bibnamefont {Leemans}},\ }\Doi {10.1038/nature02900} {\bibfield  {journal}
  {\bibinfo  {journal} {Nature (London)},\ }\textbf {\bibinfo {volume} {431}},\
  \bibinfo {pages} {538} (\bibinfo {year} {2004})}\BibitemShut {NoStop}%
\bibitem [{\citenamefont {Faure}\ \emph {et~al.}(2004)\citenamefont {Faure},
  \citenamefont {Glinec}, \citenamefont {Pukhov}, \citenamefont {Kiselev},
  \citenamefont {Gordienko}, \citenamefont {Lefebvre}, \citenamefont
  {Rousseau}, \citenamefont {Burgy},\ and\ \citenamefont
  {Malka}}]{Nature2004Faure}%
  \BibitemOpen
  \bibfield  {author} {\bibinfo {author} {\bibfnamefont {J.}~\bibnamefont
  {Faure}}, \bibinfo {author} {\bibfnamefont {Y.}~\bibnamefont {Glinec}},
  \bibinfo {author} {\bibfnamefont {A.}~\bibnamefont {Pukhov}}, \bibinfo
  {author} {\bibfnamefont {S.}~\bibnamefont {Kiselev}}, \bibinfo {author}
  {\bibfnamefont {S.}~\bibnamefont {Gordienko}}, \bibinfo {author}
  {\bibfnamefont {E.}~\bibnamefont {Lefebvre}}, \bibinfo {author}
  {\bibfnamefont {J.~P.}\ \bibnamefont {Rousseau}}, \bibinfo {author}
  {\bibfnamefont {F.}~\bibnamefont {Burgy}}, \ and\ \bibinfo {author}
  {\bibfnamefont {V.}~\bibnamefont {Malka}},\ }\Doi {10.1038/nature02963}
  {\bibfield  {journal} {\bibinfo  {journal} {Nature (London)},\ }\textbf
  {\bibinfo {volume} {431}},\ \bibinfo {pages} {541} (\bibinfo {year}
  {2004})}\BibitemShut {NoStop}%
\bibitem [{\citenamefont {Rechatin}\ \emph {et~al.}(2009)\citenamefont
  {Rechatin}, \citenamefont {Faure}, \citenamefont {Ben-Ismail}, \citenamefont
  {Lim}, \citenamefont {Fitour}, \citenamefont {Specka}, \citenamefont
  {Videau}, \citenamefont {Tafzi}, \citenamefont {Burgy},\ and\ \citenamefont
  {Malka}}]{PRL2009Rechatin1}%
  \BibitemOpen
  \bibfield  {author} {\bibinfo {author} {\bibfnamefont {C.}~\bibnamefont
  {Rechatin}}, \bibinfo {author} {\bibfnamefont {J.}~\bibnamefont {Faure}},
  \bibinfo {author} {\bibfnamefont {A.}~\bibnamefont {Ben-Ismail}}, \bibinfo
  {author} {\bibfnamefont {J.}~\bibnamefont {Lim}}, \bibinfo {author}
  {\bibfnamefont {R.}~\bibnamefont {Fitour}}, \bibinfo {author} {\bibfnamefont
  {A.}~\bibnamefont {Specka}}, \bibinfo {author} {\bibfnamefont
  {H.}~\bibnamefont {Videau}}, \bibinfo {author} {\bibfnamefont
  {A.}~\bibnamefont {Tafzi}}, \bibinfo {author} {\bibfnamefont
  {F.}~\bibnamefont {Burgy}}, \ and\ \bibinfo {author} {\bibfnamefont
  {V.}~\bibnamefont {Malka}},\ }\Doi {10.1103/PhysRevLett.102.164801}
  {\bibfield  {journal} {\bibinfo  {journal} {Phys. Rev. Lett.},\ }\textbf
  {\bibinfo {volume} {102}},\ \bibinfo {pages} {164801} (\bibinfo {year}
  {2009})}\BibitemShut {NoStop}%
\bibitem [{\citenamefont {Leemans}\ \emph {et~al.}(2006)\citenamefont
  {Leemans}, \citenamefont {Nagler}, \citenamefont {Gonsalves}, \citenamefont
  {Toth}, \citenamefont {Nakamura}, \citenamefont {Geddes}, \citenamefont
  {Esarey}, \citenamefont {Schroeder},\ and\ \citenamefont
  {Hooker}}]{NatPhys2006Leemans}%
  \BibitemOpen
  \bibfield  {author} {\bibinfo {author} {\bibfnamefont {W.~P.}\ \bibnamefont
  {Leemans}}, \bibinfo {author} {\bibfnamefont {B.}~\bibnamefont {Nagler}},
  \bibinfo {author} {\bibfnamefont {A.~J.}\ \bibnamefont {Gonsalves}}, \bibinfo
  {author} {\bibfnamefont {C.}~\bibnamefont {Toth}}, \bibinfo {author}
  {\bibfnamefont {K.}~\bibnamefont {Nakamura}}, \bibinfo {author}
  {\bibfnamefont {C.~G.~R.}\ \bibnamefont {Geddes}}, \bibinfo {author}
  {\bibfnamefont {E.}~\bibnamefont {Esarey}}, \bibinfo {author} {\bibfnamefont
  {C.~B.}\ \bibnamefont {Schroeder}}, \ and\ \bibinfo {author} {\bibfnamefont
  {S.~M.}\ \bibnamefont {Hooker}},\ }\Doi {10.1038/nphys418} {\bibfield
  {journal} {\bibinfo  {journal} {Nat. Phys.},\ }\textbf {\bibinfo {volume}
  {2}},\ \bibinfo {pages} {696} (\bibinfo {year} {2006})}\BibitemShut {NoStop}%
\bibitem [{\citenamefont {Lundh}\ \emph {et~al.}(2011)\citenamefont {Lundh},
  \citenamefont {Lim}, \citenamefont {Rechatin}, \citenamefont {Ammoura},
  \citenamefont {Ben-Isma\"il}, \citenamefont {Davoine}, \citenamefont
  {Gallot}, \citenamefont {Goddet}, \citenamefont {Lefebvre}, \citenamefont
  {Malka},\ and\ \citenamefont {Faure}}]{NatPhys2010Lundh}%
  \BibitemOpen
  \bibfield  {author} {\bibinfo {author} {\bibfnamefont {O.}~\bibnamefont
  {Lundh}}, \bibinfo {author} {\bibfnamefont {J.}~\bibnamefont {Lim}}, \bibinfo
  {author} {\bibfnamefont {C.}~\bibnamefont {Rechatin}}, \bibinfo {author}
  {\bibfnamefont {L.}~\bibnamefont {Ammoura}}, \bibinfo {author} {\bibfnamefont
  {A.}~\bibnamefont {Ben-Isma\"il}}, \bibinfo {author} {\bibfnamefont
  {X.}~\bibnamefont {Davoine}}, \bibinfo {author} {\bibfnamefont
  {G.}~\bibnamefont {Gallot}}, \bibinfo {author} {\bibfnamefont {J.-P.}\
  \bibnamefont {Goddet}}, \bibinfo {author} {\bibfnamefont {E.}~\bibnamefont
  {Lefebvre}}, \bibinfo {author} {\bibfnamefont {V.}~\bibnamefont {Malka}}, \
  and\ \bibinfo {author} {\bibfnamefont {J.}~\bibnamefont {Faure}},\ }\Doi
  {10.1038/nphys1872} {\bibfield  {journal} {\bibinfo  {journal} {Nat. Phys.},\
  }\textbf {\bibinfo {volume} {7}},\ \bibinfo {pages} {219} (\bibinfo {year}
  {2011})}\BibitemShut {NoStop}%
\bibitem [{\citenamefont {Chen}\ \emph {et~al.}(1998)\citenamefont {Chen},
  \citenamefont {Maksimchuk},\ and\ \citenamefont
  {Umstadter}}]{Nature1998Chen}%
  \BibitemOpen
  \bibfield  {author} {\bibinfo {author} {\bibfnamefont {S.~Y.}\ \bibnamefont
  {Chen}}, \bibinfo {author} {\bibfnamefont {A.}~\bibnamefont {Maksimchuk}}, \
  and\ \bibinfo {author} {\bibfnamefont {D.}~\bibnamefont {Umstadter}},\ }\Doi
  {10.1038/25303} {\bibfield  {journal} {\bibinfo  {journal} {Nature
  (London)},\ }\textbf {\bibinfo {volume} {396}},\ \bibinfo {pages} {653}
  (\bibinfo {year} {1998})}\BibitemShut {NoStop}%
\bibitem [{\citenamefont {Ta~Phuoc}\ \emph {et~al.}(2003)\citenamefont
  {Ta~Phuoc}, \citenamefont {Rousse}, \citenamefont {Pittman}, \citenamefont
  {Rousseau}, \citenamefont {Malka}, \citenamefont {Fritzler}, \citenamefont
  {Umstadter},\ and\ \citenamefont {Hulin}}]{PRL2003TaPhuoc}%
  \BibitemOpen
  \bibfield  {author} {\bibinfo {author} {\bibfnamefont {K.}~\bibnamefont
  {Ta~Phuoc}}, \bibinfo {author} {\bibfnamefont {A.}~\bibnamefont {Rousse}},
  \bibinfo {author} {\bibfnamefont {M.}~\bibnamefont {Pittman}}, \bibinfo
  {author} {\bibfnamefont {J.~P.}\ \bibnamefont {Rousseau}}, \bibinfo {author}
  {\bibfnamefont {V.}~\bibnamefont {Malka}}, \bibinfo {author} {\bibfnamefont
  {S.}~\bibnamefont {Fritzler}}, \bibinfo {author} {\bibfnamefont
  {D.}~\bibnamefont {Umstadter}}, \ and\ \bibinfo {author} {\bibfnamefont
  {D.}~\bibnamefont {Hulin}},\ }\Doi {10.1103/PhysRevLett.91.195001} {\bibfield
   {journal} {\bibinfo  {journal} {Phys. Rev. Lett.},\ }\textbf {\bibinfo
  {volume} {91}},\ \bibinfo {pages} {195001} (\bibinfo {year}
  {2003})}\BibitemShut {NoStop}%
\bibitem [{\citenamefont {Rousse}\ \emph {et~al.}(2004)\citenamefont {Rousse},
  \citenamefont {Ta~Phuoc}, \citenamefont {Shah}, \citenamefont {Pukhov},
  \citenamefont {Lefebvre}, \citenamefont {Malka}, \citenamefont {Kiselev},
  \citenamefont {Burgy}, \citenamefont {Rousseau}, \citenamefont {Umstadter},\
  and\ \citenamefont {Hulin}}]{PRL2004Rousse}%
  \BibitemOpen
  \bibfield  {author} {\bibinfo {author} {\bibfnamefont {A.}~\bibnamefont
  {Rousse}}, \bibinfo {author} {\bibfnamefont {K.}~\bibnamefont {Ta~Phuoc}},
  \bibinfo {author} {\bibfnamefont {R.}~\bibnamefont {Shah}}, \bibinfo {author}
  {\bibfnamefont {A.}~\bibnamefont {Pukhov}}, \bibinfo {author} {\bibfnamefont
  {E.}~\bibnamefont {Lefebvre}}, \bibinfo {author} {\bibfnamefont
  {V.}~\bibnamefont {Malka}}, \bibinfo {author} {\bibfnamefont
  {S.}~\bibnamefont {Kiselev}}, \bibinfo {author} {\bibfnamefont
  {F.}~\bibnamefont {Burgy}}, \bibinfo {author} {\bibfnamefont {J.~P.}\
  \bibnamefont {Rousseau}}, \bibinfo {author} {\bibfnamefont {D.}~\bibnamefont
  {Umstadter}}, \ and\ \bibinfo {author} {\bibfnamefont {D.}~\bibnamefont
  {Hulin}},\ }\Doi {10.1103/PhysRevLett.93.135005} {\bibfield  {journal}
  {\bibinfo  {journal} {Phys. Rev. Lett.},\ }\textbf {\bibinfo {volume} {93}},\
  \bibinfo {pages} {135005} (\bibinfo {year} {2004})}\BibitemShut {NoStop}%
\bibitem [{\citenamefont {Schwoerer}\ \emph {et~al.}(2006)\citenamefont
  {Schwoerer}, \citenamefont {Liesfeld}, \citenamefont {Schlenvoigt},
  \citenamefont {Amthor},\ and\ \citenamefont {Sauerbrey}}]{PRL2006Schwoerer}%
  \BibitemOpen
  \bibfield  {author} {\bibinfo {author} {\bibfnamefont {H.}~\bibnamefont
  {Schwoerer}}, \bibinfo {author} {\bibfnamefont {B.}~\bibnamefont {Liesfeld}},
  \bibinfo {author} {\bibfnamefont {H.-P.}\ \bibnamefont {Schlenvoigt}},
  \bibinfo {author} {\bibfnamefont {K.-U.}\ \bibnamefont {Amthor}}, \ and\
  \bibinfo {author} {\bibfnamefont {R.}~\bibnamefont {Sauerbrey}},\ }\Doi
  {10.1103/PhysRevLett.96.014802} {\bibfield  {journal} {\bibinfo  {journal}
  {Phys. Rev. Lett.},\ }\textbf {\bibinfo {volume} {96}},\ \bibinfo {pages}
  {014802} (\bibinfo {year} {2006})}\BibitemShut {NoStop}%
\bibitem [{\citenamefont {Schlenvoigt}\ \emph {et~al.}(2008)\citenamefont
  {Schlenvoigt}, \citenamefont {Haupt}, \citenamefont {Debus}, \citenamefont
  {Budde}, \citenamefont {Jackel}, \citenamefont {Pfotenhauer}, \citenamefont
  {Schwoerer}, \citenamefont {Rohwer}, \citenamefont {Gallacher}, \citenamefont
  {Brunetti}, \citenamefont {Shanks}, \citenamefont {Wiggins},\ and\
  \citenamefont {Jaroszynski}}]{NatPhys2008Schlenvoigt}%
  \BibitemOpen
  \bibfield  {author} {\bibinfo {author} {\bibfnamefont {H.~P.}\ \bibnamefont
  {Schlenvoigt}}, \bibinfo {author} {\bibfnamefont {K.}~\bibnamefont {Haupt}},
  \bibinfo {author} {\bibfnamefont {A.}~\bibnamefont {Debus}}, \bibinfo
  {author} {\bibfnamefont {F.}~\bibnamefont {Budde}}, \bibinfo {author}
  {\bibfnamefont {O.}~\bibnamefont {Jackel}}, \bibinfo {author} {\bibfnamefont
  {S.}~\bibnamefont {Pfotenhauer}}, \bibinfo {author} {\bibfnamefont
  {H.}~\bibnamefont {Schwoerer}}, \bibinfo {author} {\bibfnamefont
  {E.}~\bibnamefont {Rohwer}}, \bibinfo {author} {\bibfnamefont {J.~G.}\
  \bibnamefont {Gallacher}}, \bibinfo {author} {\bibfnamefont {E.}~\bibnamefont
  {Brunetti}}, \bibinfo {author} {\bibfnamefont {R.~P.}\ \bibnamefont
  {Shanks}}, \bibinfo {author} {\bibfnamefont {S.~M.}\ \bibnamefont {Wiggins}},
  \ and\ \bibinfo {author} {\bibfnamefont {D.~A.}\ \bibnamefont
  {Jaroszynski}},\ }\Doi {10.1038/nphys811} {\bibfield  {journal} {\bibinfo
  {journal} {Nat. Phys.},\ }\textbf {\bibinfo {volume} {4}},\ \bibinfo {pages}
  {130} (\bibinfo {year} {2008})}\BibitemShut {NoStop}%
\bibitem [{\citenamefont {Fuchs}\ \emph {et~al.}(2009)\citenamefont {Fuchs},
  \citenamefont {Weingartner}, \citenamefont {Popp}, \citenamefont {Major},
  \citenamefont {Becker}, \citenamefont {Osterhoff}, \citenamefont {Cortrie},
  \citenamefont {Zeitler}, \citenamefont {Horlein}, \citenamefont {Tsakiris},
  \citenamefont {Schramm}, \citenamefont {Rowlands-Rees}, \citenamefont
  {Hooker}, \citenamefont {Habs}, \citenamefont {Krausz}, \citenamefont
  {Karsch},\ and\ \citenamefont {Gruner}}]{NatPhys2009Fuchs}%
  \BibitemOpen
  \bibfield  {author} {\bibinfo {author} {\bibfnamefont {M.}~\bibnamefont
  {Fuchs}}, \bibinfo {author} {\bibfnamefont {R.}~\bibnamefont {Weingartner}},
  \bibinfo {author} {\bibfnamefont {A.}~\bibnamefont {Popp}}, \bibinfo {author}
  {\bibfnamefont {Z.}~\bibnamefont {Major}}, \bibinfo {author} {\bibfnamefont
  {S.}~\bibnamefont {Becker}}, \bibinfo {author} {\bibfnamefont
  {J.}~\bibnamefont {Osterhoff}}, \bibinfo {author} {\bibfnamefont
  {I.}~\bibnamefont {Cortrie}}, \bibinfo {author} {\bibfnamefont
  {B.}~\bibnamefont {Zeitler}}, \bibinfo {author} {\bibfnamefont
  {R.}~\bibnamefont {Horlein}}, \bibinfo {author} {\bibfnamefont {G.~D.}\
  \bibnamefont {Tsakiris}}, \bibinfo {author} {\bibfnamefont {U.}~\bibnamefont
  {Schramm}}, \bibinfo {author} {\bibfnamefont {T.~P.}\ \bibnamefont
  {Rowlands-Rees}}, \bibinfo {author} {\bibfnamefont {S.~M.}\ \bibnamefont
  {Hooker}}, \bibinfo {author} {\bibfnamefont {D.}~\bibnamefont {Habs}},
  \bibinfo {author} {\bibfnamefont {F.}~\bibnamefont {Krausz}}, \bibinfo
  {author} {\bibfnamefont {S.}~\bibnamefont {Karsch}}, \ and\ \bibinfo {author}
  {\bibfnamefont {F.}~\bibnamefont {Gruner}},\ }\Doi {10.1038/nphys1404}
  {\bibfield  {journal} {\bibinfo  {journal} {Nat. Phys.},\ }\textbf {\bibinfo
  {volume} {5}},\ \bibinfo {pages} {826} (\bibinfo {year} {2009})}\BibitemShut
  {NoStop}%
\bibitem [{\citenamefont {Joshi}\ \emph
  {et~al.}(1987){\natexlab{a}}\citenamefont {Joshi}, \citenamefont
  {Katsouleas}, \citenamefont {Dawson}, \citenamefont {Yan},\ and\
  \citenamefont {Slater}}]{PROC1987Joshi}%
  \BibitemOpen
  \bibfield  {author} {\bibinfo {author} {\bibfnamefont {C.}~\bibnamefont
  {Joshi}}, \bibinfo {author} {\bibfnamefont {T.}~\bibnamefont {Katsouleas}},
  \bibinfo {author} {\bibfnamefont {J.~M.}\ \bibnamefont {Dawson}}, \bibinfo
  {author} {\bibfnamefont {Y.~T.}\ \bibnamefont {Yan}}, \ and\ \bibinfo
  {author} {\bibfnamefont {J.~M.}\ \bibnamefont {Slater}},\ }in\ \href@noop {}
  {\emph {\bibinfo {booktitle} {Proceedings of Particle Accelerator Conf.}}},\
  \bibinfo {series and number} {\bibinfo {number} {IEEE Cat. 87c112387-9, p.
  199}}\ (\bibinfo {year} {1987})\BibitemShut {NoStop}%
\bibitem [{\citenamefont {Joshi}\ \emph
  {et~al.}(1987){\natexlab{b}}\citenamefont {Joshi}, \citenamefont
  {Katsouleas}, \citenamefont {Dawson}, \citenamefont {Yan},\ and\
  \citenamefont {Slater}}]{IEEE1987Joshi}%
  \BibitemOpen
  \bibfield  {author} {\bibinfo {author} {\bibfnamefont {C.}~\bibnamefont
  {Joshi}}, \bibinfo {author} {\bibfnamefont {T.}~\bibnamefont {Katsouleas}},
  \bibinfo {author} {\bibfnamefont {J.~M.}\ \bibnamefont {Dawson}}, \bibinfo
  {author} {\bibfnamefont {Y.~T.}\ \bibnamefont {Yan}}, \ and\ \bibinfo
  {author} {\bibfnamefont {J.~M.}\ \bibnamefont {Slater}},\ }\Doi
  {10.1109/JQE.1987.1073557} {\bibfield  {journal} {\bibinfo  {journal} {IEEE
  J. Quantum Electron.},\ }\textbf {\bibinfo {volume} {QE-23}},\ \bibinfo
  {pages} {1571} (\bibinfo {year} {1987}{\natexlab{b}})}\BibitemShut {NoStop}%
\bibitem [{\citenamefont {Clarke}(2004)}]{Clarke}%
  \BibitemOpen
  \bibfield  {author} {\bibinfo {author} {\bibfnamefont {J.~A.}\ \bibnamefont
  {Clarke}},\ }\href@noop {} {\emph {\bibinfo {title} {The Science and
  Technology of Undulators and Wigglers}}}\ (\bibinfo  {publisher} {Oxford
  University Press},\ \bibinfo {address} {Oxford},\ \bibinfo {year}
  {2004})\BibitemShut {NoStop}%
\bibitem [{\citenamefont {Williams}\ \emph
  {et~al.}(1990){\natexlab{a}}\citenamefont {Williams}, \citenamefont
  {Clayton}, \citenamefont {Joshi},\ and\ \citenamefont
  {Katsouleas}}]{RSI1990Williams}%
  \BibitemOpen
  \bibfield  {author} {\bibinfo {author} {\bibfnamefont {R.~L.}\ \bibnamefont
  {Williams}}, \bibinfo {author} {\bibfnamefont {C.~E.}\ \bibnamefont
  {Clayton}}, \bibinfo {author} {\bibfnamefont {C.}~\bibnamefont {Joshi}}, \
  and\ \bibinfo {author} {\bibfnamefont {T.}~\bibnamefont {Katsouleas}},\ }\Doi
  {10.1063/1.1141719} {\bibfield  {journal} {\bibinfo  {journal} {Rev. Sci.
  Instrum.},\ }\textbf {\bibinfo {volume} {61}},\ \bibinfo {pages} {3037}
  (\bibinfo {year} {1990}{\natexlab{a}})}\BibitemShut {NoStop}%
\bibitem [{\citenamefont {Williams}\ \emph
  {et~al.}(1990){\natexlab{b}}\citenamefont {Williams}, \citenamefont
  {Clayton}, \citenamefont {Joshi}, \citenamefont {Katsouleas}, \citenamefont
  {Mori},\ and\ \citenamefont {Slater}}]{PROC1990Williams}%
  \BibitemOpen
  \bibfield  {author} {\bibinfo {author} {\bibfnamefont {R.~L.}\ \bibnamefont
  {Williams}}, \bibinfo {author} {\bibfnamefont {C.~E.}\ \bibnamefont
  {Clayton}}, \bibinfo {author} {\bibfnamefont {C.~J.}\ \bibnamefont {Joshi}},
  \bibinfo {author} {\bibfnamefont {T.~C.}\ \bibnamefont {Katsouleas}},
  \bibinfo {author} {\bibfnamefont {W.~B.}\ \bibnamefont {Mori}}, \ and\
  \bibinfo {author} {\bibfnamefont {J.~M.}\ \bibnamefont {Slater}},\ }in\
  \href@noop {} {\emph {\bibinfo {booktitle} {Proceedings of SPIE 1227:
  Free-Electron Lasers and Applications}}},\ \bibinfo {series and number}
  {\bibinfo {number} {p. 48-59}}\ (\bibinfo {year} {1990})\BibitemShut
  {NoStop}%
\bibitem [{\citenamefont {Fedele}\ \emph {et~al.}(1990)\citenamefont {Fedele},
  \citenamefont {Milano},\ and\ \citenamefont {Vaccaro}}]{PS1990Fedele}%
  \BibitemOpen
  \bibfield  {author} {\bibinfo {author} {\bibfnamefont {R.}~\bibnamefont
  {Fedele}}, \bibinfo {author} {\bibfnamefont {G.}~\bibnamefont {Milano}}, \
  and\ \bibinfo {author} {\bibfnamefont {V.~G.}\ \bibnamefont {Vaccaro}},\
  }\Doi {10.1088/0031-8949/1990/T30/026} {\bibfield  {journal} {\bibinfo
  {journal} {Phys. Scr.},\ }\textbf {\bibinfo {volume} {1990}},\ \bibinfo
  {pages} {192} (\bibinfo {year} {1990})}\BibitemShut {NoStop}%
\bibitem [{\citenamefont {Williams}\ \emph {et~al.}(1993)\citenamefont
  {Williams}, \citenamefont {Clayton}, \citenamefont {Joshi},\ and\
  \citenamefont {Katsouleas}}]{IEEE1993Williams}%
  \BibitemOpen
  \bibfield  {author} {\bibinfo {author} {\bibfnamefont {R.}~\bibnamefont
  {Williams}}, \bibinfo {author} {\bibfnamefont {C.}~\bibnamefont {Clayton}},
  \bibinfo {author} {\bibfnamefont {C.}~\bibnamefont {Joshi}}, \ and\ \bibinfo
  {author} {\bibfnamefont {T.}~\bibnamefont {Katsouleas}},\ }\Doi
  {10.1109/27.221115} {\bibfield  {journal} {\bibinfo  {journal} {IEEE Trans.
  Plasma Sci.},\ }\textbf {\bibinfo {volume} {21}},\ \bibinfo {pages} {156 }
  (\bibinfo {year} {1993})}\BibitemShut {NoStop}%
\bibitem [{\citenamefont {Mora}\ and\ \citenamefont
  {Antonsen}(1996)}]{PRE1996Mora}%
  \BibitemOpen
  \bibfield  {author} {\bibinfo {author} {\bibfnamefont {P.}~\bibnamefont
  {Mora}}\ and\ \bibinfo {author} {\bibfnamefont {T.~M.}\ \bibnamefont
  {Antonsen}},\ }\Doi {10.1103/PhysRevE.53.R2068} {\bibfield  {journal}
  {\bibinfo  {journal} {Phys. Rev. E},\ }\textbf {\bibinfo {volume} {53}},\
  \bibinfo {pages} {R2068} (\bibinfo {year} {1996})}\BibitemShut {NoStop}%
\bibitem [{\citenamefont {Jackson}(1962)}]{Jackson}%
  \BibitemOpen
  \bibfield  {author} {\bibinfo {author} {\bibfnamefont {J.~D.}\ \bibnamefont
  {Jackson}},\ }\href@noop {} {\emph {\bibinfo {title} {Classical
  Electrodynamics}}}\ (\bibinfo  {publisher} {Wiley},\ \bibinfo {address} {New
  York},\ \bibinfo {year} {1962})\BibitemShut {NoStop}%
\bibitem [{\citenamefont {Faure}\ \emph {et~al.}(2006)\citenamefont {Faure},
  \citenamefont {Rechatin}, \citenamefont {Norlin}, \citenamefont {Lifschitz},
  \citenamefont {Glinec},\ and\ \citenamefont {Malka}}]{Nature2006Faure}%
  \BibitemOpen
  \bibfield  {author} {\bibinfo {author} {\bibfnamefont {J.}~\bibnamefont
  {Faure}}, \bibinfo {author} {\bibfnamefont {C.}~\bibnamefont {Rechatin}},
  \bibinfo {author} {\bibfnamefont {A.}~\bibnamefont {Norlin}}, \bibinfo
  {author} {\bibfnamefont {A.}~\bibnamefont {Lifschitz}}, \bibinfo {author}
  {\bibfnamefont {Y.}~\bibnamefont {Glinec}}, \ and\ \bibinfo {author}
  {\bibfnamefont {V.}~\bibnamefont {Malka}},\ }\Doi {10.1038/nature05393}
  {\bibfield  {journal} {\bibinfo  {journal} {Nature (London)},\ }\textbf
  {\bibinfo {volume} {444}},\ \bibinfo {pages} {737} (\bibinfo {year}
  {2006})}\BibitemShut {NoStop}%
\end{thebibliography}%

\end{document}